\documentclass[11pt,a4paper]{article}
\usepackage[latin1]{inputenc}
\usepackage[english]{babel}

\usepackage{amsfonts}
\usepackage{amsmath}
\usepackage{amssymb}
\usepackage{amsthm}
\usepackage{bm}
\usepackage[round]{natbib}
\usepackage{graphicx}
\usepackage{caption}
\usepackage{subcaption}
\usepackage{float}
\usepackage[parfill]{parskip}
\usepackage{epstopdf}
\usepackage{appendix}
\usepackage{booktabs}
\usepackage{verbatim}
\usepackage{url}

\newcommand{\dd}{{\mathrm d}}

\newcommand{\e}{{\rm e}}
\newcommand{\E}{{\mathbb E}}

\newcommand{\R}{{\mathbb R}}
\newcommand{\N}{{\mathbb N}}

\newcommand{\Fcal}{{\mathcal F}}
\newcommand{\Gcal}{{\mathcal G}}

\newcommand{\Scal}{{\mathcal S}}

\makeatletter
\def\thm@space@setup{%
  \thm@preskip=\parskip \thm@postskip=0pt
}
\makeatother

\newtheorem{proposition}{Proposition}[section]

\newtheorem{theorem}[proposition]{Theorem}

\newtheorem{remark}[proposition]{Remark}

\newtheorem{example}[proposition]{Example}

\title{Linear Stochastic Dividend Model}
\author{Sander Willems\footnote{EPFL and Swiss Finance Institute. E-mail: sander.willems at epfl.ch}}
\date{August 2019}

\begin{document}

\maketitle

\begin{abstract}
    In this paper we propose a new model for pricing stock and dividend derivatives. We jointly specify dynamics for the stock price and the dividend rate such that the stock price is positive and the dividend rate non-negative. In its simplest form, the model features a dividend rate that is mean-reverting around a constant fraction of the stock price. The advantage of directly specifying dynamics for the dividend rate, as opposed to the more common approach of modeling the dividend yield, is that it is easier to keep the distribution of cumulative dividends tractable. The model is non-affine but does belong to the more general class of polynomial processes, which allows us to compute all conditional moments of the stock price and the cumulative dividends explicitly. In particular, we have closed-form expressions for the prices of stock and dividend futures. Prices of stock and dividend options are accurately approximated using a moment matching technique based on the principle of maximal entropy.
\end{abstract}

\section{Introduction}
In recent years there has been an increased interest in trading dividend derivatives, in particular dividend futures. Since there also exists an active market for derivatives referencing the price of the stock paying the dividends, there is a need for derivative pricing models that can jointly price derivatives on the stock and on the dividends. Since dividend derivatives typically reference the nominal amount of dividends paid over a window of time, it seems natural to directly specify tractable dynamics for the dividend payments, or the dividend rate if dividends are paid out continuously, under a risk-neutral measure. The challenging part of this approach is to keep the stock price positive. Indeed, in absence of arbitrage and frictions such as taxes, the stock price must decrease by exactly the amount paid out as dividend, which can push the stock price in negative territory if no connection is made between the dividend and stock price dynamics. An easy solution to this problem is to model dividend yields, i.e., the fraction of the stock price that is paid out as a dividend, instead of dividends themselves. However, such a choice complicates the valuation of dividend derivatives, since their payoff now involves the product between the stock price and the dividend yield.

In this paper, we consider a stock that pays out dividends continuously\footnote{In reality, dividends are paid out discretely instead of continuously. Most liquid dividend derivatives, however, reference dividends paid by a stock index, in which case continuously paid dividends are considered an acceptable approximation.} at a rate that is stochastically varying over time. 
The dividend rate is defined as a linear function of a multivariate factor process that belongs to the class of of polynomial diffusions, see e.g.\ \cite{filipovic2016polynomial}. The drift and martingale part of the factor process and the martingale part of the stock price process are specified such that the dividend rate is non-negative and upper bounded by a constant fraction of the stock price. As a consequence, the dividend rate will go to zero as the stock price goes to zero, which guarantees the non-negativity of the stock price. The dividend yield process therefore has a zero lower bound and an upper bound that can be chosen arbitrarily large. We show that the zero boundary of the stock price is always unattainable and we provide parameter restrictions for boundary non-attainment of the factor process. The dynamics of the model becomes most intuitive with a one dimensional factor process. In this case, the dividend rate is mean-reverting around a constant fraction of the stock price. Long-term dividend futures therefore behave proportional to the stock price, while short term dividend futures have dynamics of their own. In particular, this allows to have low volatility in short-term dividend futures, while long-term dividend futures inherit the relatively high volatility of the stock price. As highlighted by \cite{buehler2018volatility}, this is a desirable feature for stochastic dividend models. Since the model belongs to the class of polynomial processes, we can compute all moments of the stock price and the (integrated) dividend rate in closed-form. In particular, this means we have closed-form prices for stock and dividend futures. Prices of stock and dividend options are efficiently approximated using a moment-matching technique based on the principle of maximal entropy, similarly as in \cite{filipovic2018term}. In a numerical study we show that option prices are approximated accurately with a small number of moments.

We show that our model does not contain a bubble in the sense that the discounted gains process is a true martingale and that the stock price is equal to the present value of all future dividends. It is important to note that the latter is not necessarily true in an arbitrage free model. Indeed, even if the discounted gains process is a true martingale, absence of arbitrage only guarantees that the present value of all future dividends is lower than or equal to the stock price. The difference between the two is equal to the present value of the stock price at an infinite time horizon, which we show to be zero in our model. This property distinguishes our model from the one of \cite{buehler2018volatility}, where the dividend rate is mean-reverting around the present value of the stock price at infinity.

The literature on dividend derivative pricing is relatively scarce.
\cite{buehler2010stochastic} proposes a model where the stock price jumps at known dividend payment dates and follows log-normal dynamics in between the payment dates. The jump amplitudes are driven by an Ornstein-Uhlenbeck process such that the stock price remains log-normally distributed and the model has closed-form prices for European call options on the stock. Dividend derivatives, however, have to be priced with Monte-Carlo simulations. In order to reconcile the high volatility in the stock price with the low volatility in dividends, they use a very negative ($-95\%$) correlation between the process driving the jump amplitudes and the stock price. An important drawback of their approach is that dividend payments are not guaranteed to be non-negative. \cite{tunaru2018dividend} uses a similar setup as \cite{buehler2010stochastic}, but uses a beta distribution for the jump amplitudes. The choice for a compactly supported jump distribution guarantees non-negative dividend payments. However, the diffusive noise of the stock has to be assumed independent of the jump amplitudes in order to have tractable expressions for dividend futures prices. Smoothing the dividends through a negative correlation between stock price and the jump amplitudes, as in \cite{buehler2010stochastic}, is therefore not possible. In a second approach, \cite{tunaru2018dividend} directly models the cumulative dividends with a logistic diffusion model. The latter has however no guarantee to be monotonically increasing, meaning that negative dividends occur frequently. Moreover, the model must be reset on an annual basis.  Option pricing is done using Monte-Carlo simulation for both methods in \cite{tunaru2018dividend}. \cite{buehler2018volatility} decomposes the stock price as the sum of a fundamental component, representing the present value of all future dividends, and a residual bubble component. The dividends are driven by a process that mean-reverts around the bubble component. The aim of this setup is to capture the stylized fact that long-term dividend futures tend to move together with the stock price, while short-term dividend futures are much less volatile. Our model, in its simplest form, shares some similarities with the approach of \cite{buehler2018volatility}. However, instead of modeling the dividends as mean-reverting around a bubble component, we choose to make them mean-revert around the stock price itself, which seems more intuitive and leads directly to the desired positive correlation between long-term dividend futures and the stock price. \cite{guennoun2017equity} consider a stochastic local volatility model for the pricing of stock and dividend derivatives. Their model guarantees a perfect fit to observed option prices, however all pricing is based on Monte-Carlo simulations.
\cite{filipovic2018term} introduce a framework based on polynomial jump-diffusions to jointly price interest rate, dividend, and stock derivatives. Our model is a special case of their general framework, but is different from the Linear Jump-Diffusion model (LJD) that was used in the numerical study of \cite{filipovic2018term}. 

The model proposed in this paper also shares some similarities with the linear hypercube model model of  \cite{ackerer2019linear} in the context of credit risk. Specifically, they specify a survival process whose drift is a linear function of a diffusive factor process with linear drift. In order for the survival process to be positive and non-increasing, they specify the factor process such that its components are all non-negative and upper bounded by the survival process. In our setup, the dividend rate is a linear function of a diffusive factor process with linear drift, which has to be specified such that the stock price is positive and the dividend rate non-negative. The stock price, whose drift is linear in the dividend rate, therefore plays a similar role as the survival process in \cite{ackerer2019linear}, but with the important difference that the stock price has a martingale part while the survival process is absolutely continuous. This martingale part requires special care and, in particular, rules out the factor process specification of the linear hypercube model of \cite{ackerer2019linear}.

The remainder of this paper is structured as follows. Section \ref{sec:model_spec} specifies the model dynamics. Section \ref{sec:derivative_pricing} discusses the pricing of stock and dividend derivatives. In Section \ref{sec:numerical_study} we calibrate a parsimonious model specification to market data.  Section \ref{sec:extensions} presents an extension of the model with jumps in the stock price. Section \ref{sec:conclusion} concludes. All proofs are collected in the appendix.

 \section{Model specification} \label{sec:model_spec}
Let $X_t$ denote the stock price process and $D_t$ the instantaneous dividend rate. Suppose for simplicity that interest rates are constant. Consider the following dynamics for $(X_t,D_t)$ under a risk-neutral measure $\mathbb{Q}$
\begin{align}
    & D_t=\mathbf{1}^\top Y_t, \label{eq:dividend}\\
    &\dd X_t =(rX_t-D_t)\,\dd t + \sigma(X_t-\frac{D_t}{a})\,\dd W_t, \label{eq:stock_dyna}\\
    & \dd Y_t=(b X_t + \beta Y_t)\,\dd t + \sqrt{X_t-\frac{D_t}{a}}\,\left[\nu_1 \sqrt{Y_{1,t}}\,\dd B_{1,t},\ldots,\nu_d \sqrt{Y_{d,t}}\,\dd B_{d,t}\right]^\top,\label{eq:factor_dyna}
\end{align}
where $r\in\R$ is the short-rate, $\sigma,a>0$, $Y_t$ is a $d$-dimensional factor process, $d\ge 1$, $b\in\R^d$, $\beta\in\R^{d\times d}$, $\nu_1,\ldots,\nu_d\ge 0$, and $(W_t,B_{1,t},\ldots,B_{d,t})$ is a $(1+d)$-dimensional standard Brownian motion. 
The following proposition provides parameter conditions such that \eqref{eq:stock_dyna}-\eqref{eq:factor_dyna} admits a unique solution taking values in
\[
E=\{(x,y)\in\R^{1+d} \colon x>0,\, y\ge 0,\,\mathbf{1}^\top y\le a x\}.
\]
\begin{proposition}
Denote by $x^-=\min(x,0)$. Suppose that
\begin{align}
&b_k+a\min_{\substack{l=1,\ldots,d\\l\neq k}}\beta_{k,l}^- \ge 0,\quad \text{for all $k\in\{1,\ldots,d\}$},\label{eq:inward_drift1}\\
    &r-a-\max_{k=1,\ldots,d}(\mathbf{1}^\top\beta)_k-\frac{\mathbf{1}^\top b}{a}\ge 0.\label{eq:inward_drift2}
\end{align}
Then for every initial value $(X_0,Y_0)\in E$ there exists a unique in law $E$-valued solution $(X_t,Y_t)$ to \eqref{eq:stock_dyna}-\eqref{eq:factor_dyna}. The solution satisfies
\begin{itemize}
    \item $Y_{k,t}>0$ for all $t\ge 0$ if $Y_{k,0}>0$ and
    \begin{equation}
        b_k+\min_{\substack{l=1,\ldots,d\\l\neq k}}\left(a\beta_{k,l}+\frac{\nu_k^2}{2}\right)^-> \frac{\nu_k^2}{2};
        \label{eq:boundary_attainment_y}
    \end{equation}
\item $a X_t > \mathbf{1}^\top Y_t$ for all $t\ge 0$ if $a X_0 > \mathbf{1}^\top Y_0$ and
\begin{equation}
r-a- \max_{k=1,\ldots,d}(\frac{\nu_k^2}{2a}+(\mathbf{1}^\top \beta)_k)-\frac{\mathbf{1}^\top b}{a}> 0.
\label{eq:boundary_attainment_x}
\end{equation}
\end{itemize}
\label{prop:state_space2}
\end{proposition}
We henceforth assume that the inequalities in \eqref{eq:inward_drift1} and \eqref{eq:inward_drift2} are satisfied and $(X_0,Y_0)\in E$. The above proposition shows in particular that we have $X_t>0$  and $D_t\ge 0$ for all $t\ge 0$. The condition in \eqref{eq:boundary_attainment_y} can be used to bound $D_t$ strictly away from zero, although this is not required from an economic point of view. The diffusive term of $X_t$ is specified such that it vanishes at the boundary $D_t=aX_t$, which is necessary to keep the process inside $E$. The condition in \eqref{eq:boundary_attainment_x} can be used to bound the stock price volatility strictly away from zero.

\begin{remark}
Remark that the more general specification $D_t=\gamma^\top Y_t$, for some $\gamma\in (0,\infty)^d$, is equivalent to the one in \eqref{eq:dividend}. Indeed, if we define $\hat{Y}_t=CY_t$ with $C=\mathrm{diag}(\gamma_1,\ldots,\gamma_d)$, then we can write $D_t=\gamma^\top Y_t=\mathbf{1}^\top \hat{Y}_t$. The dynamics of $\hat{Y}_t$ are of the same form as the dynamics of $Y_t$
\[
    \dd \hat{Y}_t=(\hat{b}X_t +\hat{\beta}\hat{Y}_t)\,\dd t+\sqrt{X_t-\frac{D_t}{a}}\,\left[\hat{\nu}_1 \sqrt{\hat{Y}_{1,t}}\,\dd B_{1,t},\ldots,\hat{\nu}_d \sqrt{\hat{Y}_{d,t}}\,\dd B_{d,t}\right]^\top,
\]
with $\hat{b}=Cb$, $\hat{\beta}=C\beta C^{-1}$, and $\hat{\nu}_k=\sqrt{\gamma_k}\nu_k$, $k=1,\ldots,d$.
\end{remark}

If we define the dividend yield $\delta_t=\frac{D_t}{X_t}$, then we have 
\[
0\le \delta_t\le a.
\]
The dividend yield is therefore bounded from above by a parameter $a>0$ of our choice. The dynamics of the log-price $x_t=\log(X_t)$ is given by
\[
\dd x_t=\left(r-\delta_t-\frac{\sigma^2}{2}(1-\frac{\delta_t}{a})^2)\right)\dd t+\sigma(1-\frac{\delta_t}{a})\,\dd W_t.
\]
The volatility of the log-price process therefore depends on the dividend yield. There is empirical evidence that dividend payout policies affect stock price volatility, see e.g.\ \cite{baskin1989dividend}, which is consistent with the dynamics of this model. In case this is not desirable, the influence of the dividend yield on the log-price volatility can be made arbitrarily small by choosing $a$ large enough.

The following proposition shows that our model does not contain a bubble in the stock price dynamics.
\begin{proposition}\label{prop:no_bubble}
The discounted gains process $G_t=\e^{-rt}X_t+\int_0^t\e^{-rs}D_s\,\dd s$ is a martingale. If $\mathbf{1}^\top b>0$, then we have for all $t\ge 0$
\begin{equation}
    \E_t\left[\int_t^\infty \e^{-r(s-t)} D_s\,\dd s\right]=X_t.\label{eq:no_bubble}
\end{equation}
\end{proposition}
Equation \eqref{eq:no_bubble} shows that the stock price is equal to the present value of all future dividends in our model. It is important to note that this is not a trivial relationship. Indeed, from no-arbitrage principles, it only follows that the present value of future dividends must be lower than or equal to the stock price, see e.g.\ \cite{filipovic2018term}. In general, if the discounted gains process is a martingale, then
\[
X_t=\E_t[\e^{-r(T-t)}X_T]+\E_t\left[\int_t^T \e^{-r(s-t)} D_s\,\dd s\right],\quad T\ge t.
\]
A positive difference between the stock price and the present value of future dividends can be interpreted as the present value of a terminal payment at an infinite time horizon, which is difficult to reconcile with standard economic theory. Proposition \ref{prop:no_bubble} shows that, in our model, we have $\displaystyle\lim_{T\to \infty}\E_t[\e^{-r(T-t)}X_T]=0$ if $\mathbf{1}^\top b>0$. The derivation of this result relies on the linear drift structure of $(X_t,Y_t)$ and the geometry of $E$, which are a key ingredients of our model. In Example \ref{ex:bubble} in the next section, we illustrate a parameterization where the assumption $\mathbf{1}^\top b>0$ is violated and \eqref{eq:no_bubble} does not hold.

\begin{remark}
The processes $X_t$ and $Y_t$ have zero quadratic covariation. Note that this does not mean that dividends are independent of the stock price, since $X_t$ still enters in the drift and diffusion function of $Y_t$. The dynamics of $X_t$ can be generalized to allow for non-zero quadratic covariation with $Y_t$ as follows
\[
\dd X_t =(rX_t-D_t)\,\dd t + \sigma(X_t-\frac{D_t}{a})\,\dd W_t
+\sqrt{X_t-\frac{D_t}{a}}\sum_{k=1}^d \eta_k \sqrt{Y_{k,t}}\dd B_{k,t},
\]
for some parameters $\eta_k\in \R$, $k=1,\ldots,d$. All the results in the paper are easily adjusted to accommodate this generalization.
\end{remark}

\subsection{Single factor model}
For $d=1$, we obtain the following model dynamics
\begin{align}
    &\dd X_t =(rX_t-D_t)\,\dd t +\sigma(X_t-\frac{D_t}{a})\,\dd W_t, \label{eq:stock_dyna_1d}\\
    &\dd D_t=(b X_t+\beta D_t)\,\dd t + \nu_1 \sqrt{D_t(X_t-\frac{D_t}{a})}\,\dd B_t. \label{eq:factor_dyna_1d}
\end{align}
If $\beta<0$, then $D_t$ is mean-reverting around $-\frac{b}{\beta} X_t$, with an upper bound of $aX_t$. The inward pointing drift conditions \eqref{eq:inward_drift1} and \eqref{eq:inward_drift2} become
\begin{equation}
0\le b\le a(r-a-\beta).
\label{eq:inward_drift_1d}
\end{equation}
Boundary non-attainment is satisfied if $0<D_0<aX_0$ and 
\[
\frac{\nu_1^2}{2}< b < a(r-a-\beta)-\frac{\nu_1^2}{2}.
\]
The dividend yield becomes an autonomous diffusion with the following dynamics
\[
\dd \delta_t=\left(b+(\beta-r)\delta_t+\delta_t^2+\sigma^2\delta_t(1-\frac{\delta_t}{a})^2\right)\dd t -\sigma \delta_t(1-\frac{\delta_t}{a})\dd W_t+\nu_1\sqrt{\delta_t(1-\frac{\delta_t}{a})}\dd B_t.
\]
Remark that the dividend yield process is not a polynomial diffusion, due to the terms $\delta^2$ and $\delta^3$ in the drift, and $\delta^3$ and $\delta^4$ in the diffusion function. However, since $\delta_t$ is typically in the order of percentage points, higher powers of $\delta$ contribute relatively little to the dynamics. In particular, the dividend yield $\delta_t$  has approximately a linear drift $b+(\beta-r+\sigma^2)\delta_t$.

We end this section with an example where the assumption in Proposition \ref{prop:no_bubble} is violated.
\begin{example} \label{ex:bubble}
If $b=0$, then we have $\E_t[D_T]=\e^{\beta (T-t)}D_t$ for all $T\ge t$, and  \eqref{eq:inward_drift_1d} becomes $r-\beta\ge a>0$. The present value of future dividends is
\[
    \E_t\left[\int_t^\infty \e^{-r(s-t)}D_s\,\dd s\right]=\int_t^\infty \e^{(\beta-r)(s-t)}\,\dd s\, D_t=\frac{D_t}{r-\beta}.
\]
Using $aX_t \ge D_t$ and $r-\beta\ge a$ we obtain
\[
 \E_t\left[\int_t^\infty \e^{-r(s-t)}D_s\,\dd s\right] \le \frac{a}{r-\beta}X_t \le X_t.
\]
The present value of future dividends is therefore lower than or equal to the stock price, as required by absence of arbitrage. If $aX_t>D_t$ or $r-\beta >a$, then we have an example where the present value of future dividends is strictly below the stock price, i.e., $\E_t[\e^{-r(T-t)}X_T]$ does not go to zero as $T\to \infty$.
\end{example}
The above example shows that the presence of $X_t$ in the drift of $D_t$ is crucial for the stock price to be equal to the present value of future dividends.

\section{Derivative pricing}\label{sec:derivative_pricing}
In this section we show how to compute prices of derivatives referencing the stock price and/or the dividends paid over some time interval. 
\subsection{Moments of stock price and cumulative dividends}
Define the cumulative dividend process as
\[
C_t=\int_0^t D_s\,\dd s,\quad t\ge 0,
\]
which represents all the dividends paid out over a time interval $[0,t]$. In contrast to the instantaneous dividend rate $D_t$, the cumulative dividend $C_t$ is observable in practice. The process $(C_t,X_t,Y_t)$ is jointly a polynomial diffusion, so we are able to compute all conditional moments in closed form, see e.g.\ \cite{filipovic2016polynomial} for details.  Let $\mathrm{Pol}_n$ denote the set of polynomials $p\colon \R^{2+d}\to\R$ with $1\le \deg(p)\le n$. Applying the infinitesimal generator $\Gcal$ of $(C_t,X_t,Y_t)$ to a twice differentiable function $f(c,x,y)$ gives
\[
\Gcal f=(\mathbf{1}^\top y,rx-\mathbf{1}^\top y,(bx+\beta y)^\top) \nabla f
+\frac{1}{2}\sigma^2(x-\frac{1^\top y}{a})^2f_{xx}
+\frac{1}{2}\sum_{k=1}^d \nu_k^2 y_k(x-\frac{\mathbf{1}^\top y}{a})f_{y_ky_k},
\]
where the subscripts of $f$ denote partial derivatives, $\nabla f$ the gradient of $f$, and we have omitted the function arguments for brevity. It is easily verified that $\Gcal \mathrm{Pol}_n \subseteq \mathrm{Pol}_n$ for any $n\in\N$. Therefore, if we fix a vector of polynomial basis functions $H_n=(h_1,\ldots,h_{N_n})^\top$ for $\mathrm{Pol}_n$, with $N_n=\dim (\mathrm{Pol}_n)$, then we can find a unique matrix $G_n$ such that for all $ (c,x,y^\top)^\top\in\R^{2+d}$
\[
\Gcal H_n(c,x,y)=G_n H_n(c,x,y).
\]
By definition of the infinitesimal generator, we obtain the following moment formula
\begin{equation}
    \E_t[H_n(C_T,X_T,Y_T)]=\e^{G_n(T-t)}H_n(C_t,X_t,Y_t),\quad \forall T\ge t.
    \label{eq:moment_formula}
\end{equation}
In particular, we can compute all the $\Fcal_t$-conditional mixed moments of $(C_T,X_T)$ in closed form for all $T\ge t$.

\begin{remark}
If one is only interested in the moments of $X_T$, then there is no need to augment the state with $C_t$, since $(X_t,Y_t)$ is already a polynomial diffusion on its own.
\end{remark}

\subsection{Linear derivatives}
For $n=1$, we can without loss of generality choose the basis $H_1(c,x,y)=(c,x,y^\top)^\top$. The matrix $G_1$ then becomes 
\[
G_1=
\begin{pmatrix}
0 & 0 &\mathbf{1}^\top\\
0& r & -\mathbf{1}^\top\\
0 & b & \beta
\end{pmatrix}.
\]
The most actively traded linear derivatives are stock futures and dividend futures. Stock futures settle on the stock price at some terminal date $T$ and dividend futures settle on the dividends paid in a time interval $[T_0,T_1]$. The moment formula \eqref{eq:moment_formula} can be used to compute prices of stock futures and dividend futures. Indeed, futures prices are given by the risk-neutral expectation of the terminal settlement price because of continuous marking-to-market, so we get
\begin{align}
&\E_t[X_T]=\mathbf{e}_2^\top\e^{G_1(T-t)}H_1(C_t,X_t,Y_t),\label{eq:stock_futures_price}\\
&\E_t[C_{T_1}-C_{T_0}]=\mathbf{e}_1^\top\left(\e^{G_1(T_1-t)}-\e^{G_1(T_0-t)}\right)H_1(C_t,X_t,Y_t),\label{eq:dividend_futures_prices}
\end{align}
where $\mathbf{e}_k$ denotes the $k$-th canonical basis vector in $\R^{2+d}$, $T\ge t$, and $T_1\ge T_0\ge t$. In case the reference period of the dividend futures has already started, i.e., $0\le T_0\le t\le T_1$, we get
\begin{equation}
    \E_t[C_{T_1}-C_{T_0}]=\mathbf{e}_1^\top\e^{G_1(T_1-t)}H_1(C_t,X_t,Y_t) - C_{T_0}.
\end{equation}
Without loss of generality we can assume that $T_0=0$, in which case $C_{T_0}=0$ and $C_t$ is the amount of dividends already paid.

Remark that the volatility parameters $\sigma$ and $\nu_1,\ldots,\nu_d$ do not enter into the prices of dividend futures, which is a consequence of the linear drift structure of $Y_t$. This allows us, for example, to calibrate $b$ and $\beta$ to dividend futures first, and subsequently use $\sigma$ and $\nu_1,\ldots,\nu_d$ to calibrate non-linear derivatives such as stock and dividend options. The parameter $a$ also does not appear in the prices of dividend futures, however it should be noted that the value of $a$ affects the values that $b$ and $\beta$ are allowed to take, because of the inequalities \eqref{eq:inward_drift1} and \eqref{eq:inward_drift2} that we assume to be true.

\subsection{Non-linear derivatives}\label{sec:non-lin}
Consider a derivative on the stock price with discounted payoff at time $T$ given by $F(X_T)$, for some function $F$. In absence of arbitrage, its price at time $t\le T$ is given by
\[
\pi_t=\E_t[F(X_T)].
\]
The probability density function of $X_T$ is not known explicitly, so we cannot compute $\pi_t$ by direct integration in general. We do however know all the moments of $X_T$ through the moment formula \eqref{eq:moment_formula}. In particular, if $F$ is a polynomial, then we can compute $\pi_t$ explicitly. If $F$ is not a polynomial, we approximate $\pi_t$ using the available stock price moments and the principle of maximum entropy, similarly as in \cite{filipovic2018term}. Specifically, denote by $M_n=\E_t[X_T^n]$, $n=1,\ldots, N$, the first $N\ge 1$ moments of the stock price. We now look for a probability density function $f$ which has the same first $N$ moments as $X_T$ and has maximal entropy:
\begin{equation}
\def\arraystretch{2}
\begin{array}{ccc}
 \displaystyle \max_{f}&-\displaystyle\int_0^\infty f(x)\ln f(x)\,\dd x&\\
 \mathrm{s.t.}&\displaystyle\int_0^\infty x^n f(x)\,\dd x=M_n,&\quad n=0,\ldots,N,
\end{array}
\label{eq:entropy_optim}
\end{equation}
where we set $M_0=1$ so that the density integrates to one. \cite{jaynes1982rationale} motivates such a choice by noting that maximizing entropy incorporates the least amount of prior information in the distribution, other than the imposed moment constraints. In this sense it is maximally noncommittal with respect to unknown information about the distribution. Straightforward functional variation with respect to $f$ gives the following unique solution $f^{(N)}$ to the optimization problem in \eqref{eq:entropy_optim}
\begin{align*}
f^{(N)}(x)=\exp\left(-\sum_{n=0}^N \lambda_n x^n\right),
\end{align*}
where the Lagrange multipliers $\lambda_0,\ldots,\lambda_N$ have to be solved numerically from the moment constraints. Finally, we approximate $\pi_t$ by numerically computing the integral
\[
\pi^{(N)}=\int_0^\infty F(x)f^{(N)}(x)\,\dd x.
\]

We can use exactly the same approach to price dividend derivatives with discounted payoff at time $T_1$ given by $P(C_{T_1}-C_{T_0})$, for some function $P$. All we need are the moments of $C_{T_1}-C_{T_0}$, which can be computed explicitly using the law of iterated expectations and the moment formula \eqref{eq:moment_formula} as follows
\begin{align*}
    \E_t[(C_{T_1}-C_{T_0})^n]&=\sum_{k=0}^n {n\choose k}\E_t[(-C_{T_0})^{n-k}\E_{T_0}[C_{T_1}^k]]\\
    &=\sum_{k=0}^n {n\choose k}\E_t[(-C_{T_0})^{n-k}w^\top_k \e^{G_k(T_1-T_0)}\tilde{H}_k(C_{T_0},X_{T_0},Y_{T_0})]\\
    &=\sum_{k=0}^n {n\choose k}v^\top_k \e^{G_n(T_0-t)}H_n(C_{t},X_{t},Y_{t}),
\end{align*}
where $v_k$ and $w_k$ are the unique vectors satisfying $w_k^\top H_k(c,x,y)=c^k$ and $v_k^\top H_n(c,x,y)=(-c)^{n-k}w^\top_k \e^{G_k(T_1-T_0)}H_k(c,x,y)$.

\section{Numerical study} \label{sec:numerical_study}
As an example, we calibrate the single factor model \eqref{eq:stock_dyna_1d}--\eqref{eq:factor_dyna_1d} using a snapshot of real market data on 21/12/2015. The stock in the calibration exercise is the Euro Stoxx 50 index, the leading blue-chip stock index in the Eurozone. The Euro Stoxx 50 index is well suited for calibrating our model since it has a liquid dividend derivatives market associated with it. The Euro Stoxx 50 dividend futures contracts are exchange traded on Eurex and
reference the sum of the declared ordinary gross cash dividends (or cash-equivalent) on index constituents that go ex-dividend during a given calendar year, divided by
the index divisor on the ex-dividend day. There are always ten adjacent annual contracts available for trading, with maturities every third Friday of December. We use all ten contracts in the calibration. Euro Stoxx 50 dividend options are also exchange traded on Eurex. They are options on the futures contracts, where the maturity of the option coincides with the maturity of the futures contract, which makes them effectively options on the dividends realized in a calendar year. In the calibration, we use the Black implied volatility of the option on the first dividend futures contract with at-the-money strike (i.e., strike equal to the dividend futures price). We also use the Black-Scholes implied volatility of the option on the Euro Stoxx 50 index level with maturity in three months and at-the-money strike. The prices of the dividend futures and the implied volatility of the index and dividend option are shown in the second column of Table \ref{tab:prices_errors}.  Remark that the implied volatility of the dividend option is substantially lower ($\approx 5\%$) than the implied volatility of the index option ($\approx 23\%$). This immediately shows that models with a constant dividend yield are not appropriate to price dividend derivatives, since they produce dividend payments that are roughly as volatile as the stock price itself.

\begin{table}
    \centering
    \begin{tabular}{lcccc}\toprule
        &  &\multicolumn{3}{c}{Absolute errors}\\ \cline{3-5}
       &Data & $a=0.1$ & $a=0.2$ & $a=0.3$\\ \midrule 
DF1 & 115.3 & 0.183 & 0.183 & 0.183 \\ 
DF2 & 108.7 & 0.492 & 0.492 & 0.492 \\ 
DF3 & 105.5 & 1.452 & 1.452 & 1.451 \\ 
DF4 & 100.1 & 0.344 & 0.344 & 0.344 \\ 
DF5 & 95.7 & 0.399 & 0.399 & 0.399 \\ 
DF6 & 92.0 & 0.918 & 0.918 & 0.918 \\ 
DF7 & 89.6 & 0.497 & 0.497 & 0.497 \\ 
DF8 & 87.2 & 0.350 & 0.349 & 0.349 \\ 
DF9 & 84.8 & 0.414 & 0.413 & 0.413 \\ 
DF10 & 84.6 & 1.558 & 1.558 & 1.558 \\ 
IV stock&0.2295 & 4.095e-07 & 9.381e-07 & 9.87e-08 \\ 
IV dividend&0.0491 & 2.001e-07 & 6.854e-07 & 9.395e-07 \\ \bottomrule 
    \end{tabular}
    \caption{The second column shows market data as of 21/12/2015. DF$k$ represents the dividend futures contract with expiry on the third Friday of December $(2015+k)$. IV stock is the Black-Scholes implied volatility of the stock option. IV dividend is the Black implied volatility of the dividend option. All data comes from Bloomberg. The last three columns show the absolute errors of the calibrated models.}
    \label{tab:prices_errors}
\end{table}

\begin{figure}
    \centering
    \includegraphics[width=0.75\textwidth]{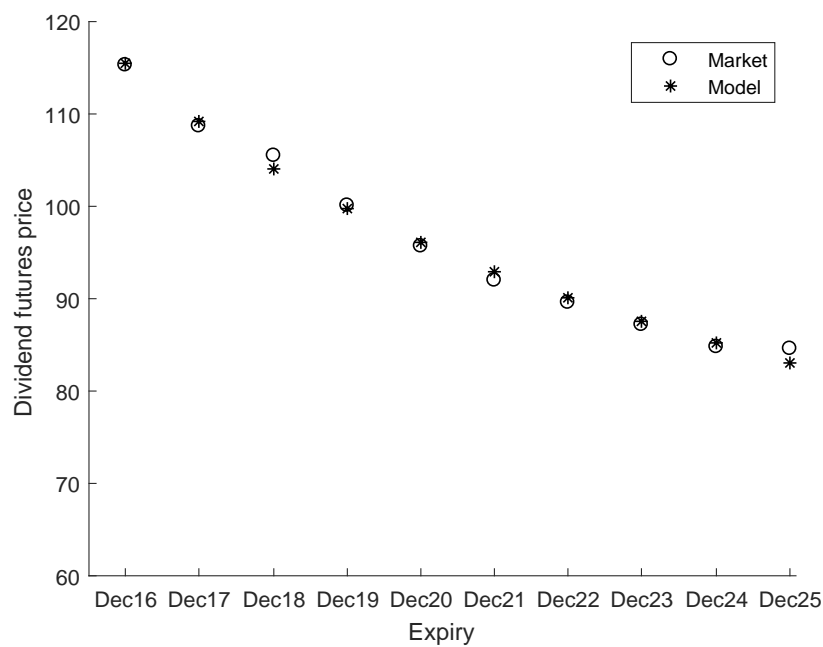}
    \caption{Market prices and model implied prices of dividend futures. The model implied prices are calculated using the parameter in Table \ref{tab:parameter_calibration} with $a=0.2$.}
    \label{fig:div_fut}
\end{figure}

We fix $r=0.01$ and $a\in\{0.1,0.2,0.3\}$. By fixing $a$, the parameter constraint in \eqref{eq:inward_drift_1d} becomes a linear inequality in the free parameters $b$ and $\beta$, which most optimization routines can easily deal with. In our model, $a$ determines the upper bound for the dividend yield process $\delta_t$. In Figure \ref{fig:div_fut} we plot a proxy of the (unobservable) dividend yield $\delta_t$ over time, which we calculate by dividing the price of the first to expire dividend futures contract (which has a time to maturity varying between 1 day and 1 year) by the index level. We observe that between 2010 and 2016, the dividend yield proxy moves roughly between 3\% and 6\%, well below the three values that we consider for $a$. 

\begin{figure}
    \centering
    \includegraphics[width=0.75\textwidth]{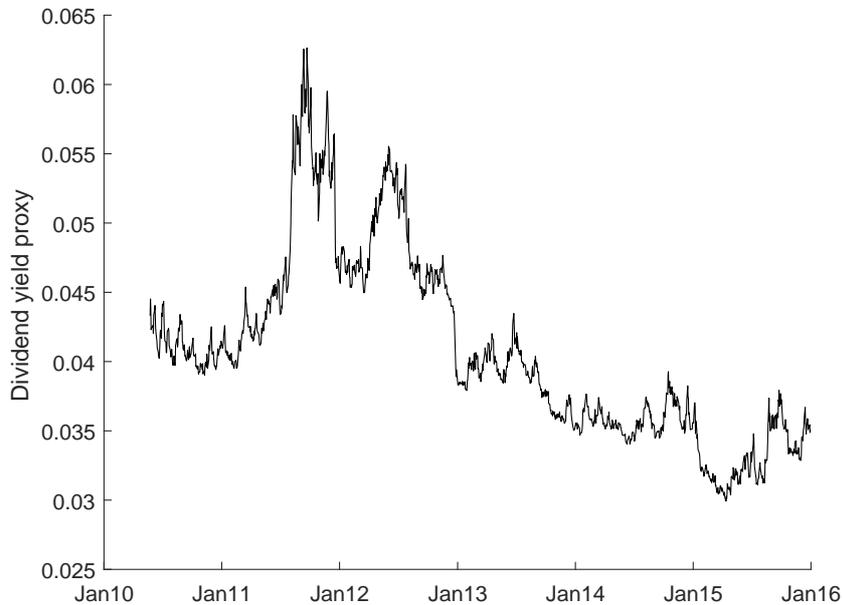}
    \caption{This figure plots the historical dividend yield, which we proxy by the price of the first to expire dividend futures contract divided by the index level.}
    \label{fig:div_yield}
\end{figure}

We use $N=6$ moments to compute the dividend and stock option prices using the maximal entropy method described in Section \ref{sec:non-lin}.  We use the gradient-free Nelder-Mead simplex optimization algorithm to find the optimal parameters $b,\beta,\sigma,\nu_1$, and $D_0$. The calibrated parameters are shown in Table \ref{tab:parameter_calibration} and the absolute pricing errors are shown in the last three columns of Table \ref{tab:prices_errors}. The calibrated values of $b$, $\beta$, and $D_0$ are almost identical for different values of $a$. This is not surprising, since these parameters mainly control the term structure of dividend futures prices, and $a$ does not enter in the pricing formula \eqref{eq:dividend_futures_prices} for the dividend futures.\footnote{Indirectly, the dividend futures prices are to some extent affected by $a$ through the inequality \eqref{eq:inward_drift_1d} that has to be satisfied.} However, from \eqref{eq:stock_dyna_1d} and \eqref{eq:factor_dyna_1d} it is clear that $a$ has an impact on the volatility of the stock price and the dividend rate. Indeed, if $a$ increases, all else being equal, then the volatility of the stock price and the dividend rate increases. To offset this effect, the calibrated parameters of $\sigma$ and $\nu_1$ are smaller for larger $a$. From the absolute errors in Table \ref{tab:prices_errors}, we can see that the choice of $a$ does not really matter for the quality of the calibration, since the absolute pricing errors are almost identical. The maximal relative error in the dividend futures contracts is less than $2\%$, which is a remarkably good fit for a single factor model. Figure \ref{fig:div_fut} visualizes the good fit of the calibrated model with the dividend futures term structure. The option prices are matched perfectly. This is a consequence of the fact that the dividend futures prices do not depend on the martingale part of $X_t$ and $D_t$. The parameters $\sigma$ and $\nu_1$ therefore remain free to calibrate to the dividend and stock option.

\begin{figure}
    \centering
    \includegraphics[width=0.75\textwidth]{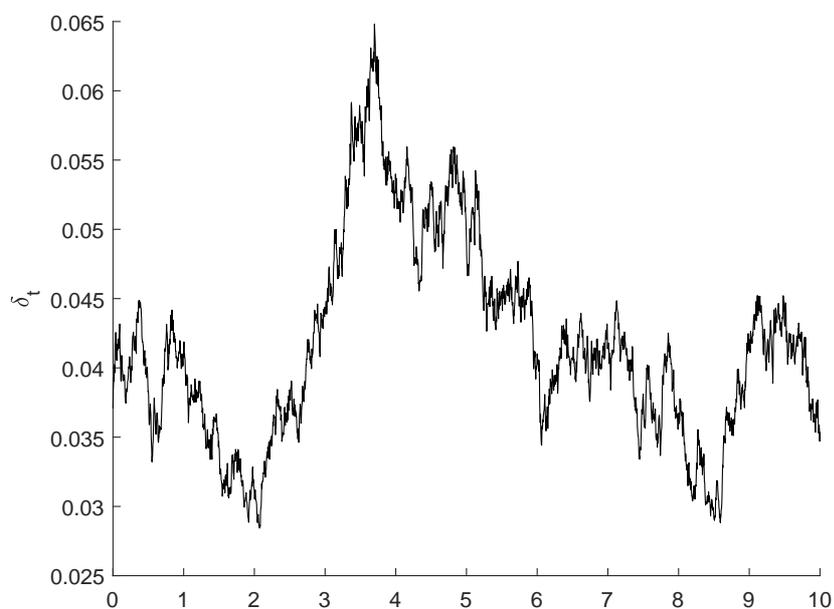}
    \caption{Simulation the dividend yield process $\delta_t$ over ten years with daily discretization. The model parameters are those in Table \ref{tab:parameter_calibration} with $a=0.2$.}
    \label{fig:div_yield_sim}
\end{figure}

Figure \ref{fig:div_yield_sim} plots a simulation of the dividend yield process $\delta_t$ over a ten years horizon with daily discretization. We use the model parameters from Table \ref{tab:parameter_calibration} with $a=0.2$, however the plot looks identical when using the calibrated parameters with $a=0.1$ or $a=0.3$. The process is roughly mean-reverting around $b/(r-\beta-\sigma^2)=3.61\%$, which is the mean-reversion level of $\delta_t$ when ignoring the higher order terms in the drift. Remark that the range of values that $\delta_t$ takes in the simulation is very similar to the range of values in Figure \ref{fig:div_yield}.

\begin{figure}
    \centering
    \begin{subfigure}[b]{0.45\textwidth}
        \includegraphics[width=\textwidth]{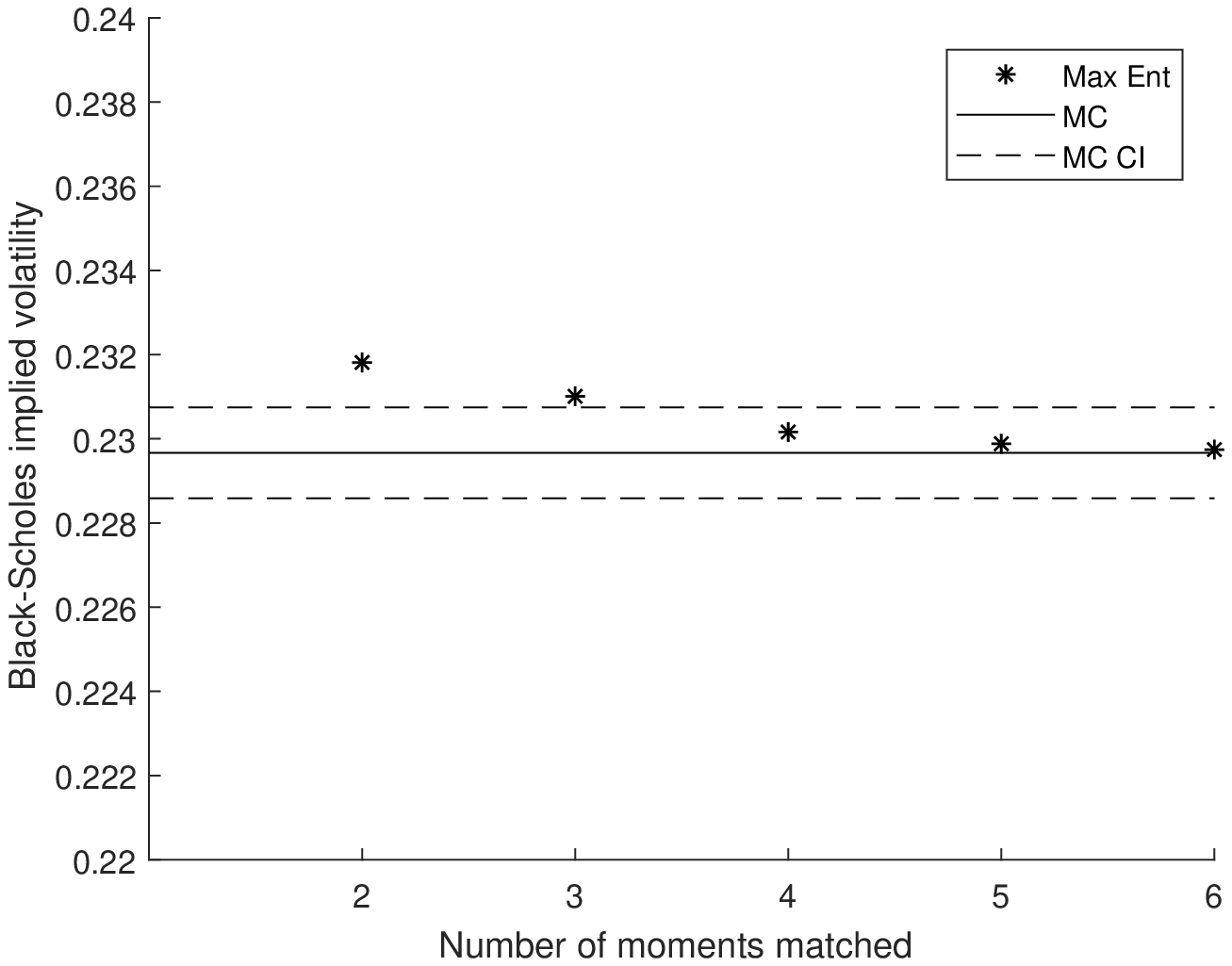}
        \caption{Stock option}
        \label{fig:mom_conv_stock}
    \end{subfigure}
    ~ 
    \begin{subfigure}[b]{0.45\textwidth}
        \includegraphics[width=\textwidth]{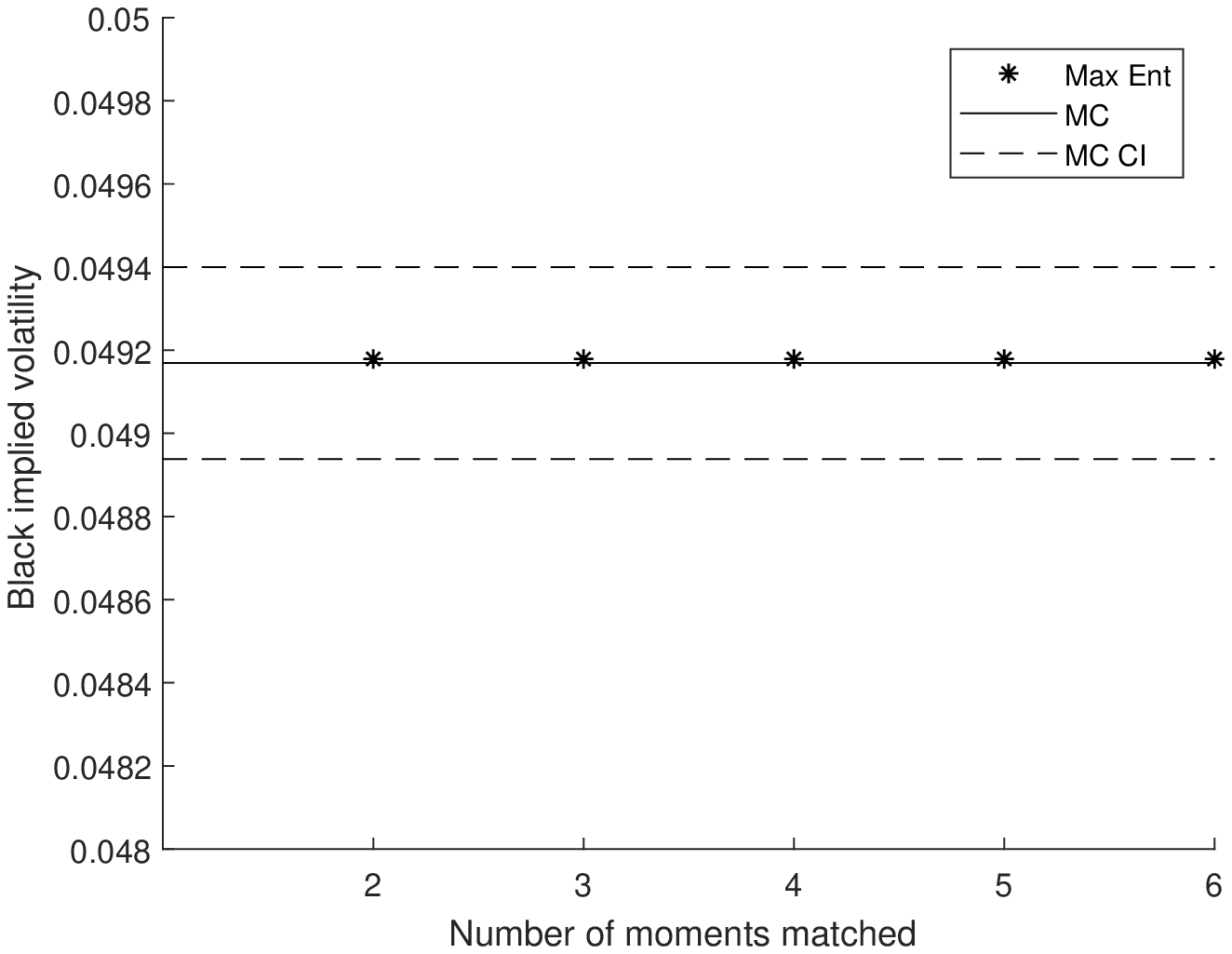}
        \caption{Dividend option}
        \label{fig:mom_conv_div}
    \end{subfigure}
    \caption{Option price approximations for varying number of moments $N$. The solid lines represents the Monte-Carlo estimates and the dashed lines represent the corresponding 95\% confidence intervals.}
    \label{fig:mom_conv}
\end{figure}

Figure \ref{fig:mom_conv} shows the option price approximation as a function of the number of moments $N$. As a benchmark, we run a Monte-Carlo simulation with daily time steps and $10^5$ sample paths. In order to reduce the variance of the Monte-Carlo estimator, we use a degree one polynomial in the underlying as a control variate, where we determine the coefficients of the polynomial through a least squares regression of the simulated payoff paths on the simulated polynomial. The solid line shows the Monte-Carlo estimator and the dashed lines are the corresponding $95\%$ confidence intervals. For the stock option, the maximal entropy approximation with four moments is already within the confidence bands and the one with six moments is exactly equal to the Monte-Carlo estimator. For the dividend option, using only two moments already provides a very accurate option price approximation. The dividend option price is easier to approximate because the volatility of the dividend rate is much lower than the volatility of the stock price.

\begin{table}
    \centering
    \begin{tabular}{c|ccccc} \toprule
        $a$ &  $b$ & $\beta$ & $\sigma$ & $\nu_1$ & $D_0$  \\ \hline 
         0.1 & 0.0103 & -0.3440 & 0.3621 & 0.0220 & 0.0371 \\
         0.2 & 0.0103 & -0.3439 & 0.2813 & 0.0194 & 0.0371 \\
         0.3 & 0.0103 & -0.3439 & 0.2614 & 0.0187 & 0.0371 \\ \bottomrule 
    \end{tabular}
    \caption{Calibrated model parameters for three different values of $a$.}
    \label{tab:parameter_calibration}
\end{table}

\section{Extending the model with jumps}\label{sec:extensions}
We can enrich the model dynamics by adding jumps to $X_t$ as follows
\begin{align}
    &\dd X_t =(rX_t-D_t)\,\dd t +(X_{t-}-\frac{D_{t-}}{a})\,(\sigma \dd W_t + \dd J_t), \label{eq:stock_dyna_jump}
\end{align}
where $D_t$ is the same as before and $J_t$ is a compensated compound Poisson process with arrival intensity $\lambda\ge 0$ and with a jump distribution $F(\dd z)$ that is assumed to have moments in closed-form of all orders and a support $\Scal\subseteq (-1,\infty)$. Remark that $D_t$ is still a continuous process, so that $D_{t-}=D_t$. Let $\tau$ denote a jump time of $J_t$ and suppose that $(X_{\tau-},Y_{\tau-})\in E$. From the assumption on the support of $F$, we have 
\[
X_\tau=X_{\tau-}+\Delta X_\tau =X_{\tau-}+(X_{\tau-}-\frac{D_{\tau-}}{a})\Delta J_\tau\ge \frac{D_{\tau-}}{a},
\]
where equality only holds if $a X_{\tau -}=D_{\tau -}$, in which case $\Delta X_\tau =0$. Therefore, the results in Proposition \ref{prop:state_space2} remain valid since $X_t$ behaves as in \eqref{eq:stock_dyna} in between jump times, we have $(X_{\tau},Y_{\tau})\in E$ so the process cannot jump outside of $E$, and $aX_{\tau}>D_{\tau}$ if $aX_{\tau-}>D_{\tau-}$ so jumps to the boundary are not possible.

If we denote by $\Gcal_J$ the the infinitesimal generator of $(C_t,X_t,Y_t)$ in the case with jumps, then we get for a twice differential function $f$
\begin{align*}
\Gcal_J f= \Gcal f+\lambda \int_\Scal f\left(c,x+xz-\frac{\mathbf{1}^\top y}{a}z,y\right)-f-\left(x-\frac{\mathbf{1}^\top y}{a}\right)zf_x \,F(\dd z),
\end{align*}
where we assume that $f$ is such that the integral is finite and we have again omitted the function arguments for brevity, except in the first term of the integrand. Since the amplitudes of the jumps in $X_t$ depend linearly on $X_t$ and $Y_t$, it follows immediately that $\Gcal_J \mathrm{Pol_n}\subseteq \mathrm{Pol_n}$. Therefore, $(C_t,X_t,Y_t)$ belongs to the class of polynomial jump-diffusions and we can compute all conditional moments in closed form.

Since $X_t$ enters in the dynamics of $Y_t$, the jumps also indirectly impact the dynamics of $D_t$. The magnitude of the effect of a jump in $X_t$ on the drift of $D_t$ is determined by $b$. A stylized fact of index options and index dividend options is that both have a negative skew in implied volatilities. Choosing a distribution $F$ with a sufficiently negative mean produces a negative skew in implied volatilities for both stock and dividend options. We leave a calibration to option skews for future work.

\begin{remark}
It is possible to introduce jumps in $Y_t$ as well, although one should be careful with simultaneous jumps where $D_t$ jumps up and $X_t$ jumps down in order to avoid jumping out of $E$. We do not consider this extension in this paper.
\end{remark}

\section{Conclusion}\label{sec:conclusion}
We have introduced a model for jointly pricing stock and dividend derivatives. The novelty of our approach lies in the fact that we directly model the dividend rate while guaranteeing a positive stock price. This is accomplished by upper bounding the dividend rate by a constant fraction of the stock price, so that the dividend rate goes to zero as the stock price approaches zero. The model belongs to the class of polynomial diffusions, which leads to closed-form prices for stock and dividend futures, and efficient approximations for stock and dividend options. We have calibrated a single factor model to data on dividend futures and at-the-money stock and dividend options. Future research includes calibrating the model to stock and dividend options with a range of strikes using the extension with jumps outlined in Section \ref{sec:extensions}, as well as extending our framework to discrete dividend payments.

\clearpage 
\appendix
\section{Proofs}
\subsection{Proof of Proposition \ref{prop:state_space2}}
Existence of an $\R^{1+d}$-valued solution follows from \cite[Theorem IV.2.4]{ikeda1981stochastic}, since the drift and dispersion coefficient of $(X_t,Y_t)$ satisfy a linear growth condition. It remains to show that a solution starting in $E$ also stays in $E$.

Denote by $\mu\colon E\to\R$ and $\Sigma\colon E \to\R^{d\times d}$ respectively the drift and dispersion function of $Y_t$, i.e.
\[
\dd Y_t=\mu(X_t,Y_t)\,\dd t+\Sigma(X_t,Y_t)\,\dd B_t.
\]
 We need to verify that $\mu_k(x,y)\ge 0$ for $(x,y)\in E$ with $y_k=0$, so that the drift pushes $Y_{k,t}$ away from the zero boundary again. Using the fact that $0\le y_k \le ax$ for all $(x,y)\in E$, we have for all $(x,y)\in E$ with $y_k=0$ that
\begin{align*}
    \mu_k(x,y)&=b_k x +\sum_{l\neq k}\beta_{k,l}y_{l}\\
    &\ge b_k x+\min_{l\neq k}\beta_{k,l}^-\sum_{l\neq k}y_l \\
    &\ge (b_k+a\min_{l\neq k}\beta_{k,l}^-)x\ge 0
\end{align*}
The above inequality, together with $\Sigma_{k,l}(x,y)=0$, $l=1,\ldots,d$, for $(x,y)\in E$ with $y_k=0$, shows that $Y_{k,t} \ge 0$ for all $t\ge 0$ and all $k=1,\ldots,d$. Indeed, $Y_{k,t}$ starts in $E$ and has an inward pointing drift and vanishing diffusion at the boundary. Using the same argument for $a X_t - \mathbf{1}^\top Y_t$ instead of $Y_{k,t}$, it follows that $a X_t \ge \mathbf{1}^\top Y_t$ for all $t\ge 0$. As a consequence we also have $X_t\ge 0$.

In order to prove the non-attainment of the zero lower boundary of $X_t$, we use a stochastic comparison argument. Define the process $Z_t=-\log X_t$ if $X_t>0$ and $Z_t=\infty$ if $X_t=0$. Define the process $\tilde{Z}_t$ through the following SDE
\[
\dd \tilde{Z}_t=(a-r+\frac{1}{2}\sigma^2)\,\dd t-\sigma\,\dd W_t,\quad \tilde{Z}_0=Z_0.
\]
From Theorem 1.3 in \cite{hajek1985mean} we get for all $c\in\R$ and $t>0$ 
\[
P(Z_t\ge c)\le 2P(\tilde{Z}_t\ge c).
\]
Since $\displaystyle\lim_{c\to\infty}P(\tilde{Z}_t\ge c)=0$, we have $\displaystyle \lim_{c\to\infty}P(Z_t\ge c)=0$ and therefore $X_t>0$ a.s.

We use Theorem 5.7(i) in \cite{filipovic2016polynomial}, which we restate below for completeness, to study boundary attainment of $Y_t$.
\begin{theorem}[Theorem 5.7(i) \cite{filipovic2016polynomial}]
Denote by $\Gcal$ the infinitesimal generator and by $m(x,y)$ the diffusion function of $(X_t,Y_t)$. Let $p(x,y)$ be a polynomial and let $h(x,y)$ be a vector of polynomials such that $m(x,y) \nabla p(x,y)=h(x,y)p(x,y)$ for all $(x,y)\in\R^{1+d}$. If there exists a neighborhood $U$ of $E\cap \{p=0\}$ such that for all $(x,y)\in E\cap U$
\begin{equation}
2\Gcal p(x,y) - h(x,y)^\top \nabla p(x,y) \ge 0,
\label{eq:boundary_attainment}
\end{equation}
then $p(X_t,Y_t)>0$ for all $t>0$.
\label{thm:boundary_attainment}
\end{theorem}
First, we derive conditions such that $Y_{k,t}>0$. For $p(x,y)=y_k$, we have 
\[h(x,y)=(0,\ldots,\nu_k^2(x-1^\top y/a),\ldots,0)^\top,\]
with the non-zero element in the $(k+1)$-th component. For some $\epsilon>0$, consider the following neighborhood of $E\cap \{p=0\}$
\[
U=\{(x,y)\in\R^{1+d}\colon |y_k| \le \epsilon\}.
\]
For $(x,y)\in E\cap U=\{(x,y)\in\R^{1+d}\colon x>0,y_k \le \epsilon , ax\ge \mathbf{1}^\top y, y\ge 0 \}$ we have
\begin{align}
     2\Gcal p(x,y)-h(x,y)^\top\nabla p(x,y)
     &=2(b_k x +\sum_{l=1}^d\beta_{k,l}y_l)-(x-\mathbf{1}^\top y/a) \nu_k^2\nonumber \\
     &\ge 2(b_k x +\sum_{l\neq k}\beta_{k,l}y_l )-(x-\sum_{l\neq k}y_l/a) \nu_k^2 +\min(2\beta_{k,k}+\nu_k^2/a,0)\epsilon\nonumber \\
     &= (2b_k-\nu_k^2)x+\sum_{l\neq k}(2\beta_{k,l} +\nu_k^2/a)y_l+\min(2\beta_{k,k}+\nu_k^2/a,0)\epsilon\nonumber \\
     &\ge (2b_k-\nu_k^2+\min_{l\neq d}(2a\beta_{k,l}+\nu_k^2)^-)x+\min(2\beta_{k,k}+\nu_k^2/a,0)\epsilon.
     \label{eq:pos_Y_proof}
\end{align}
If $2\beta_{k,k}+\nu_k^2/a\ge 0$, then \eqref{eq:pos_Y_proof} is non-negative if $2b_k-\nu_k^2+\min_{l\neq k} (2a\beta_{k,l} +\nu_k^2)^-\ge 0$. If $2\beta_{k,k}+\nu_k^2/a<0$, then we can always find an $\epsilon>0$ such that \eqref{eq:pos_Y_proof} is non-negative if $2b_k-\nu_k^2+\min_{l\neq k} (2a\beta_{k,l} +\nu_k^2)^-> 0$.

Finally, we derive conditions such that $a X_t >\mathbf{1}^\top Y_t $. For $p(x,y)=ax-\mathbf{1}^\top y$ we have
\[
h(x,y)=(\sigma^2(x-\mathbf{1}^\top y/a),-y_1\nu_1^2/a,\ldots,-y_d\nu_d^2/a)^\top.
\]
For some $\epsilon>0$, consider the following neighborhood of $E\cap \{p=0\}$
\[
U=\{(x,y)\in\R^{1+d}\colon |ax-\mathbf{1}^\top y| \le \epsilon\}.
\]
For $(x,y)\in E\cap U=\{(x,y)\in\R^{1+d}\colon x>0, 0\le ax-\mathbf{1}^\top y \le \epsilon,y\ge 0\}$ we have
\begin{align*}
       &2\Gcal p(x,y)-h(x,y)^\top\nabla p(x,y)\\
       &=2a(rx-\mathbf{1}^\top y)-2\mathbf{1}^\top (bx+\beta y) -\sigma^2 a(x-\mathbf{1}^\top y/a)-y_1\nu_1^2/a-\cdots -y_d\nu_d^2/a\\
       &=(2ar-a\sigma^2-2\mathbf{1}^\top b)x-((2a-\sigma^2)\mathbf{1}^\top+(\nu_1^2,\ldots,\nu_d^2)/a+2\mathbf{1}^\top \beta)y\\
       &\ge (2ar-a\sigma^2-2\mathbf{1}^\top b)x-\max_{k=1,\ldots,d}(2a-\sigma^2+\nu_k^2/a+2\mathbf{1}^\top \beta_k)\mathbf{1}^\top y\\
       &\ge (2a(r-a)-2\mathbf{1}^\top b -a \max_{k=1,\ldots,d}(\nu_k^2/a+2\mathbf{1}^\top \beta_k))x +\epsilon \min(0,2a-\sigma^2+\max_{k=1,\ldots,d}(\nu_k^2/a+2\mathbf{1}^\top \beta_k)),
\end{align*}
where the last line follows from $ax-\epsilon \le \mathbf{1}^\top y\le ax$.
If $2a-\sigma^2+\max_{k=1,\ldots,d}(\nu_k^2/a+2\mathbf{1}^\top \beta_k) \ge 0$, then $2\Gcal p-h^\top\nabla p \ge 0$ on $E\cap U$ if
\[
2a(r-a)-2\mathbf{1}^\top b -a \max_{k=1,\ldots,d}(\nu_k^2/a+2\mathbf{1}^\top \beta_k)\ge 0.
\]
If $2a-\sigma^2+\max_{k=1,\ldots,d}(\nu_k^2/a+2\mathbf{1}^\top \beta_k)<0$, then we can always find an $\epsilon>0$ such that $2\Gcal p-h^\top\nabla p \ge 0$ on $E\cap U$ if
\[
2a(r-a)-2\mathbf{1}^\top b -a \max_{k=1,\ldots,d}(\nu_k^2/a+2\mathbf{1}^\top \beta_k)> 0.
\]

For uniqueness in law of the solution $(X_t,Y_t)$, note that $\frac{Y_t}{X_t}$ is an autonomous diffusion with $0\le \frac{Y_t}{X_t}\le a$ for all $t\ge 0$. A straightforward application of It\^o's lemma shows that the process $(\log(X_t),\frac{Y_t}{X_t})$ has a uniformly bounded drift and diffusion function, so that uniqueness in law for $(\log(X_t),\frac{Y_t}{X_t})$, and therefore for $(X_t,Y_t)$, follows from \cite[Theorem IV.3.3]{ikeda1981stochastic}.

\subsection{Proof of Proposition \ref{prop:no_bubble}}
To proof that $G_t$ is a martingale, we can use Novikov's condition. An application of It\^o's lemma gives
\[
\dd G_t=\sigma\e^{-rt}(X_t-\frac{D_t}{a})\,\dd W_t,
\]
which shows that $G_t$ is a local martingale. Since the volatility of $\log(G_t)$ is uniformly bounded,
\[
\left|\frac{\sigma\e^{-rt}(X_t-\frac{D_t}{a})}{\e^{-rt}X_t+\int_0^t\e^{-rs}D_s\,\dd s}\right|\le \sigma,
\]
Novikov's condition is trivially satisfied and we conclude that $G_t$ is a martingale.
\begin{remark}
The process $G_t$ represents the discounted value of a trading strategy of a long position in the stock and investing all the dividends in the risk-free account. Alternatively, we could also re-invest all the dividends in the stock itself. This strategy has a discounted value process $G^\ast_t=\e^{-rt +\int_0^t\delta_s\,\dd s}X_t$, which is again a martingale by Novikov's condition.
\end{remark}

Next, we show that the present value of all future dividends is equal to the stock price. Define $\tilde{X}_t=\e^{-rt}X_t$ and $\tilde{Y}_t=\e^{-rt}Y_t$. The dynamics of $\tilde{X}_t$ and $\tilde{Y}_t$ becomes
\begin{align*}
&\dd \tilde{X}_s=-\mathbf{1}^\top \tilde{Y}_s\,\dd s+\cdots \dd W_s,\\
&\dd \tilde{Y}_s=(b\tilde{X}_s+(\beta-r\mathrm{Id})\tilde{Y}_s)\,\dd s + \cdots \dd B_s.
\end{align*}
Taking conditional expectations and denoting $f(s)=\E_t[\tilde{X}_s]$ and $g(s)=\E_t[\tilde{Y}_s]$, $s\ge t$, gives the following linear first order ODE
\[
\begin{pmatrix}
f'\\ g'
\end{pmatrix}
=
\begin{pmatrix}
0& -\mathbf{1}^\top\\ b & \beta-r\mathrm{Id}
\end{pmatrix}
\begin{pmatrix}
f\\ g
\end{pmatrix}.
\]
Using the properties of $E$, we have that $f(s)>0$, $g(s)\ge 0$, and $\mathbf{1}^\top g(s) \le a f(s)$ for all $s\ge t$. In particular, we have that $f$ is a non-increasing function and hence $f$ and $g$ are uniformly bounded
\[
0< f(s) \le f(t),\quad 0\le g(s)\le a f(t),\quad \forall s\ge t.
\]
Moreover, all derivatives of $f$ and $g$ are uniformly bounded as well since
\[
\begin{pmatrix}
f^{(n)}\\ g^{(n)}
\end{pmatrix}
=
\begin{pmatrix}
0& -\mathbf{1}^\top\\ b & \beta-r\mathrm{Id}
\end{pmatrix}^n
\begin{pmatrix}
f\\ g
\end{pmatrix},
\]
for all $n\in\N$. Since $f$ is a non-increasing positive function, we have $\displaystyle\lim_{s\to\infty}f(s)=\xi\in[0,f(t)]$. Since $f''$ is bounded, $f'$ must be uniformly continuous. By Barbalat's lemma (see e.g., Lemma 8.2 in \cite{khalil2002nonlinear}) we therefore have that $\displaystyle\lim_{s\to\infty}f'(s)=0$. Since $f'(s)=-\mathbf{1}^\top g(s)$ and $g(s)\ge 0$, we also have $\displaystyle\lim_{s\to\infty}g(s)=0$ componentwise. Taking the limit of $f''$ gives
\[
\lim_{s\to\infty} f''(s)=
\lim_{s\to\infty}-\mathbf{1}^\top g'(s)=-\mathbf{1}^\top b\lim_{s\to\infty}f(s)-\mathbf{1}^\top (\beta-r\mathrm{Id})\lim_{s\to\infty}g(s)=-\mathbf{1}^\top b\xi.
\]
Since $f'''$ is bounded, $f''$ must be uniformly continuous, and by Barbalat's lemma we have $\displaystyle\lim_{s\to\infty}f''(s)=0$. Since $\mathbf{1}^\top b>0$ by assumption, we must have $\xi=0$. This concludes the proof since
\[
0=\lim_{s\to\infty}f(s)=\lim_{s\to\infty}\E_t[\tilde{X}_s]=\tilde{X}_t-\E_t\left[\int_t^\infty \e^{-r u}D_u\,\dd u\right].
\]

\bibliography{references}
\bibliographystyle{chicago}
\end{document}